\documentclass[12pt]{article}
\usepackage[utf8]{inputenc}
\usepackage{graphicx}
\usepackage[numbers]{natbib}
\usepackage{color}
\usepackage{amsmath}
\usepackage{amssymb}
\usepackage{amsfonts}
 \usepackage{hyperref}
\usepackage[algo2e]{algorithm2e} 
\usepackage{algorithm}
\usepackage{verbatim}
\usepackage[left=1in,right=1in,top=1in,bottom=1in]{geometry}

\title{Statistical Network Analysis: Past, Present, and Future
\footnote{This invited article is intended as a book chapter for the volume ``Frontiers of Statistics and Data Science'' edited by Subhashis Ghoshal and Anindya Roy for the International Indian Statistical Association Series on Statistics and Data Science, published by Springer. This review article covers the material from the short course titled ``Statistical Network Analysis: Past, Present, and Future'' taught by the author at the Annual Conference of the International Indian Statistical Association, June 6-10, 2023, at Golden, Colorado.}}
\author{Srijan Sengupta}
\date{North Carolina State University}

\begin{document}

\maketitle

\begin{abstract}
\noindent
We live in a highly interconnected world where many physical, social, biological, and technological systems consist of agents or entities interacting with each other. Examples include a virus being transmitted over social contact networks, global trade between countries, and the human brain. Any such system can be represented as a network, by denoting the agents/entities as vertices, and the interactions between them as edges. This makes networks an important and ubiquitous type of data spanning a remarkable variety of complex systems. 
It is therefore very important to have mathematically rigorous and practically useful methods for statistical analysis of networks. However, the structure and configuration of networks are quite different from that of traditional forms of statistical data, which means that new statistical methodology is needed for realistic modeling and reliable inference for network data. Fittingly, the last two decades have seen a remarkable surge in research aimed at developing statistical methodology for network data. 
This article provides a brief overview of this rapidly evolving field of statistics, which encompasses statistical models, algorithms, and inferential methods for analyzing data in the form of networks. 
Particular emphasis is given to connecting the historical developments in network science to today's statistical network analysis, and outlining important new areas for future research.

\end{abstract}

\section{Introduction}
Many complex systems in today's world consist, at an abstract level, of \textit{agents} who \textit{interact} with one another.
This general agent-interaction framework describes many interesting and important systems, such as the human brain \citep{bassett2017small}, online communities \citep{adamic2005political,larsen2022statistical}, and power grids \citep{watts1998collective}, to name a few.
Networks  provide a convenient and unified way of representing such systems arising from diverse applications.

Statistical network analysis is a key pillar of network science.
Quoting from the founding editorial of the journal {Network Science}: 
 ``\textit{Statistics is often defined as the study of data, involving anything from its collection, preparation, and management to its exploration, analysis, and presentation.
 In this view, 
our definition of network science delineates a subarea of statistics
concerned with data of a peculiar format} \citep{brandes2013network}.'' 
Fittingly, network data inference has recently developed into a major methodological research area, with applications spanning a range of scientific disciplines.


This article provides a brief overview of this important and rapidly evolving area of statistics.
At a conservative estimate, probably hundreds of papers have been written in the statistical literature on network data, and thousands of papers have been written on the broader topic of network science.
It is clearly infeasible and counter-productive to attempt to provide a coherent summary of this vast literature in a single article.
Instead, the goal of this article is to tell the \textit{story} of this fascinating research area,  with an eye on connecting the historical developments in network science to today's statistical network analysis and outlining some important new areas for future research.
The aim of this article is to be \textit{comprehensible} rather than comprehensive, and several prominent topics are therefore excluded, some of which are mentioned in the discussion.
For readers looking for a more comprehensive review of network science, an excellent reference is \cite{newman2010networks}.
For readers looking for statistically oriented overviews of network data inference, two excellent references are \cite{goldenberg2010survey} and \cite{kolaczyk2009statistical}.

The rest of the article is structured as follows.
Section 2 describes some key early developments and historical perspectives that led to the development of network science. 
Section 3 provides a brief outline of the current landscape of the methodological literature on network data.
Section 4 focuses on two emerging topics of future interest: scalability and network properties.
Section 5 concludes the article with a brief discussion.

\section{Past: key early developments }

Scientific research on  networks has a long and rich legacy dating back almost three centuries to Leonhard Euler’s celebrated analysis of the ``Seven Bridges of Königsberg" problem \citep{euler1741solutio}.
In this chapter, we outline some of the key early developments in network science, focusing on the topics that are particularly relevant to today's statistical network analysis.

\subsection{The Seven Bridges of Königsberg}
In the city of Königsberg, Prussia (now Kaliningrad, Russia), seven bridges spanned the Pregel River, connecting various parts of the city. The citizens posed a challenge: was it possible to take a walk through the city, crossing each bridge exactly once and returning to the starting point? Leonhard Euler approached this problem not by drawing maps or planning routes but by abstracting the problem into nodes (landmasses) and edges (bridges). In doing so, Euler essentially laid the groundwork for graph theory almost three centuries back \citep{euler1741solutio}. 

Euler's key observation was this: during any walk in the graph, the number of times one enters a non-terminal vertex equals the number of times one leaves it. 
Now, if every edge is traversed exactly once, each non-terminal vertex must have an even number of edges.
Through his analysis, he concluded that such a walk was impossible given the structure of the Seven Bridges of Königsberg network.
This analysis approach introduced the key concepts of paths and cycles that are central to network science. 

He further proposed a theorem, now known as Euler's theorem, which states that a graph has an Eulerian circuit (a cycle that visits every edge exactly once and returns to the starting vertex) if and only if every vertex has an even degree. 
On a historical note, while Euler discussed the necessary conditions for the existence of Eulerian paths and circuits, it was Carl Hierholzer who provided a detailed proof and algorithm for finding Eulerian paths and circuits in 1873 \citep{hierholzer1873moglichkeit}.

\subsection{Sociograms}
The first half of the twentieth century saw an important milestone in the development of modern network science: the use of \textit{sociograms} for visual and mathematical representation of networks.
A sociogram is a graphical representation where individual entities (vertices) are connected by relationships (edges). Moreno and Jennings \citep{moreno1934shall} utilized this concept within a classroom setting, aiming to capture the interpersonal dynamics by asking a straightforward question: who preferred to sit next to whom? Each student's choices were then represented visually to display the relational dynamics within the group.

In their analysis, Moreno and Jennings observed that interpersonal relationships within the classroom displayed certain evolving patterns. For instance, over time, the frequency of mutual choices between boys and girls decreased, potentially indicating the emergence of specific social norms. Additionally, the formation and dissolution of closely interconnected clusters of students were observed, underscoring the evolving nature of social relationships.
Their study identified network motifs such as cliques—groups with dense interconnections—and stars, which indicated individuals frequently chosen by others but whose choosers weren't necessarily interconnected themselves. The study of these motifs later became instrumental in the field of social network analysis and more broadly, network science.

Sociograms emphasized the importance of graphical representation in deciphering complex relational data. This approach facilitated the comprehension of higher-order connections and paved the way for future methodologies in network science.
Sociograms can also be considered the precursors to adjacency matrices widely used in today's statistical network analysis.

\subsection{Spectral Graph Theory}
Spectral Graph Theory, the study of network properties from the perspective of eigenvalues, eigenvectors, and characteristic polynomials of network matrices such as adjacency matrices or Laplacian matrices,
developed as a sub-discipline of random matrix theory in the 1950s and the 1960s \citep{wigner1958distribution}.
We start by defining some key related concepts.
A \textbf{walk} of length $k$ between nodes is defined as a sequence of $k$ edges connecting them. A \textbf{path} is a unique walk without repetitive vertices, whereas a \textbf{closed walk} returns to the initiating vertex. The \textbf{adjacency matrix} $A$ of dimension $n \times n$ is such that $A(i,j) = 1$ if nodes $i$ and $j$ are directly connected, and $0$ otherwise. A pivotal result from SGT is that the count of closed walks of length $k$ in a graph corresponds to $\sum_{j=1}^n \lambda_j^k$, where $\lambda_1, \ldots, \lambda_n$ are the eigenvalues of $A$ \citep{van2010graph}.
This result provides a natural and elegant connection between network paths and eigenvalues.

Next, we briefly describe three key results/concepts in spectral graph theory that are especially relevant to statistical network analysis.

\subsubsection{Cheeger's Inequality and Fiedler Vector}
Cheeger's inequality characterizes the relationship between the edge expansion of a graph and the second smallest eigenvalue of the graph Laplacian matrix  \citep{chung2005laplacians,van2010graph}. 
Consider a graph $G$ whose set of vertices is given by $V$ and set of edges is given by $E$.
The Laplacian matrix, denoted by $L$, is defined as
\[ L = D - A \]
where $D$ is the diagonal matrix of vertex degrees. 
The edge expansion \( h(G) \) of $G$ is defined for a subset of vertices \( S \subset V \) with \( 0 < |S| \leq \frac{|V|}{2} \) as
\[
h(S) = \frac{|\partial(S)|}{|S|}
\]
where \( |\partial(S)| \) is the number of edges with one end in \( S \) and the other in \( V \backslash S \). The edge expansion of the graph \( h(G) \) is the minimum of \( h(S) \) over all such subsets \( S \).
Cheeger's inequality states that
\[ 2h(G) \geq \lambda_2 \geq \frac{h(G)^2}{2 \Delta(G)} \]
with $h(G)$ representing edge expansion, $\Delta(G)$ denoting the maximum degree for the nodes in $G$, and $\lambda_2$ denoting the second smallest eigenvalue of the graph Laplacian matrix.

The Fiedler vector is the eigenvector associated with the second smallest eigenvalue of the Laplacian matrix.
The Fiedler vector characterizes the connectivity of a graph, partitioning it into two segments based on the sign of its components. This partition finds the minimal edge expansion cut, effectively identifying the graph's densely connected regions, and providing the conceptual bedrock of the popular spectral clustering approach for graph clustering \citep{von2007tutorial,rohe2011spectral}
These results have several connections and applications to modern statistical network analysis.
The Fiedler vector is instrumental in spectral graph partitioning and aids in visualizing the graph structure.
Furthermore, Cheeger's inequality provides insights into 
 the connectivity of a graph. It helps measure 
how easily information or influence can spread within a network, making it relevant in analyzing the robustness and efficiency of networks.

\subsubsection{Epidemic Threshold}
Consider the following question: a pathogen (e.g., virus) is introduced into an entirely uninfected population connected via a social contact network.
Will the transmission die out or lead to an epidemic?
Specifically, consider the SIR epidemiological model: the population is divided into three compartments based on the disease status of individuals: susceptible (S), infected (I), and recovered (R).
All individuals start in the susceptible compartment, meaning they are susceptible to the disease. When a susceptible individual comes into contact with an infected individual, there is a probability of infection. Once infected, an individual moves to the infected compartment. Infected individuals can then transmit the disease to susceptible individuals.
The spread of the disease is driven by two parameters: $\beta$, the infection rate per effective contact and $\mu$, the recovery rate.
Transmissibility is defined as $\gamma = \beta/\mu$, and 
critical transmissibility is defined as $\gamma_c$ such that $\gamma > \gamma_c$ leads to an epidemic.

A classical result from network epidemiology is that the epidemic threshold for an arbitrary static network is given by  
\begin{equation*}
    \gamma_c = \frac{1}{\lambda[{A}]},
\end{equation*}
where $\lambda[{A}]$ is the spectral radius of the adjacency matrix ${A}$ \citep{Prakash2010,leitch2019toward}.
This result is elegant because it separates out the boundary condition of the epidemic threshold on two sides: the left-hand side being purely a function of the pathogen and the right-hand side being purely a function of the graph.
Therefore, the eigenvalues of the adjacency matrix play a crucial role in determining the epidemic threshold. For example, the impact of movement restrictions on reducing the risk of an epidemic can be measured by the change in $\lambda[\mathbf{A}]$.

\subsubsection{Co-spectral Graphs}
Co-spectral graphs, also known as isospectral graphs, are graphs that share the same set of eigenvalues.  Two graphs, G and H, are said to be co-spectral if the spectra of their respective adjacency matrices or Laplacian matrices are exactly the same.
Given an undirected graph G with n vertices, the spectrum of G refers to the set of eigenvalues of its adjacency matrix or its Laplacian matrix. Two graphs, G and H, are said to be co-spectral if the spectra of their respective adjacency matrices or Laplacian matrices are exactly the same.

Co-spectral graphs are interesting because they are highly similar without necessarily being isomorphic.
Note that co-spectral graphs share any property defined by eigenvalues, e.g., epidemic threshold, Cheeger constant, number of closed paths of any order, etc. 
Co-spectral graphs have implications in various areas, including cryptography, network analysis, and theoretical computer science. For example, in the field of graph isomorphism testing, which aims to determine whether two graphs are structurally identical, co-spectral graphs provide examples where the eigenvalues alone cannot distinguish between the graphs \citep{van2010graph}.

\subsection{Random graphs}
The study of networks as random objects saw significant strides in the 1950s and 1960s, with the most notable being the so-called Erdős–Rényi model which was independently introduced by Paul Erdős and Alfréd Rényi, and Edgar Gilbert \citep{erdHos1959random, erdHos1960evolution, gilbert1959random}.
The Erdős–Rényi random graph model \( G(n, p) \) consists of a graph with \( n \) vertices where each pair of vertices is connected by an edge with probability \( p \), independently from every other pair.

One of the most fascinating and well-studied properties of this model is the emergence of the ``giant component.'' As the edge probability increases, there exists a critical threshold beyond which a large connected component containing a positive fraction of the vertices, referred to as the giant component, suddenly appears, marking a phase transition in the graph's structure.
The emergence of the giant component occurs as follows:
\begin{itemize}
  \item For \( np(n) < 1 \), the graph almost surely consists of small, disconnected components, and the size \( |C_{max}| \) of the largest component \( C_{max} \) is of order \( \log(n) \).
  \item At \( np(n) = 1 \), there is a phase transition.
  \item For \( np(n) > 1 \), there exists a giant component \( C_{max} \) such that \( |C_{max}| \) is of order \( n \), meaning that \( \frac{|C_{max}|}{n} \) converges in probability to a positive number less than 1 as \( n \) goes to infinity.
\end{itemize}

The phase transition at \( np(n) = 1 \) marks the critical point where the structure of the graph changes dramatically, resulting in the emergence of a giant component.
This probabilistic analysis of networks as random graphs paved the way for statistical modeling and inference of network data.



\section{Present: Statistical Inference of Network Data}

Today, {network science} is recognized as an academic field in its own right, 
with dedicated professional societies, journals, 
conferences, and academic departments \citep{NAP11516}.
This was spurred, in part, by a number of influential and roughly contemporaneous papers published between 1998 and 2001, which unveiled the remarkable fact that networks from widely different fields possess common properties \citep{watts1998collective, newman2001structure,barabasi1999emergence, borgatti2000models}.
This included the small world property, core-periphery structure, and community structure, to name a few.
These papers and subsequent research convinced scientists from a wide range of disciplines that there is a {\textit{science of networks}}, a scientific commonality, that lies at the foundation of their research disciplines, and helped launch the field of network science.

\begin{itemize}
    \item \textbf{{The small-world property}:}
The small-world property consists of ``{segregation}" of vertices into small tightly knit groups that leads to high local clustering in an otherwise sparse network, while at the same time the network has a small average path length that ``{integrates}'' the network \citep{watts1998collective, lovekar2021testing}.
The {global clustering coefficient} \( C \) for a graph \( G \) with \( n \) vertices is defined as the average of the local clustering coefficients of all the nodes in the graph:
\[ C = \frac{1}{n} \sum_{i=1}^{n} C_i \]
where \( C_i \) is the local clustering coefficient of vertex \( i \), given by:
\[ C_i = \frac{2|E_i|}{k_i(k_i - 1)} \]
In this, \( |E_i| \) represents the number of edges between the neighbors of \( i \) and \( k_i \) is the degree of \( i \).
The {average path length} \( L \) in the network is the average number of steps along the shortest paths for all possible pairs of network nodes:
\[ L = \frac{1}{n(n-1)} \sum_{i \neq j} d(i, j) \]
where \( d(i, j) \) is the shortest path length between vertices \( i \) and \( j \).
 The commonly accepted {metric for quantifying the small-world property is through the \textit{small-world coefficient}} defined as
 $$\sigma = \frac{{C}/C_R}{{L}/L_R},$$
 where ${C}$ and ${L}$ are observed global clustering coefficient and average path length respectively, while $C_R$ and $L_R$ are the expected values of the same quantities in a Erd\"{o}s-Ren\'{y}i random graph (ER) of equivalent density.
 Over the last two decades,  ``small-world'' networks have been observed in a wide variety of disciplines, such as neuroscience \citep{bassett2017small},  scientific collaboration \cite{newman2001structure}, and air transportation networks \cite{guimera2005worldwide}.

\item{\textbf{Community Structure:}}
Community structure is a mesoscale phenomenon where the network consists of tightly-knit groups of nodes, called communities, with denser connections amongst themselves than connections with the rest of the network (Figure \ref{fig:comm_cp}, left).
The vertices in the same groups or communities often
display similar behavior, while vertices from different communities behave differently.
Identifying this network structure, called community detection,
is an important problem in network science \citep{newman2004finding,bickel2009nonparametric,yanchenko2021generalized}.
Community detection has a wide range of scientific applications, ranging from social networks where they might represent social groups or affiliations, to biological and biomedical networks where they can signify functional units or complexes \citep{fortunato2010community, jonsson2006cluster, komolafe2022scalable, boxley2022using}.

\item {\textbf{Core-periphery structure:}}
A network is said to exhibit {\it core-periphery (CP) structure} (Figure \ref{fig:comm_cp}, right). when the network can be split into two parts: a {\it core} which is densely connected with itself and the periphery, and a {\it periphery} which is loosely connected with itself \citep{borgatti2000models,YanSen2023}. This differs from the traditional assortative community structure \citep[e.g.,][]{newman2004finding} where communities are densely connected with themselves but loosely connected with each other.
CP structures have been observed in real-world networks across many different domains, e,g,  global trade networks, where large economy nations form the core \citep{krugman1996self, magone2016core}, and airport networks where airline hubs fly all across the country and comprise the core whereas regional airports are mainly connected to large airports and not to other small airports, a hallmark of peripheral nodes \citep{lordan2017analyzing, lordan2019core}. In both of these domains, the node clusters (core and periphery) share some underlying properties, so identifying these groups yields information about the overall network structure.
\begin{figure}
    \centering
        \includegraphics[width=0.3\textwidth, trim={2.5cm 2.5cm 2cm 2cm},clip]{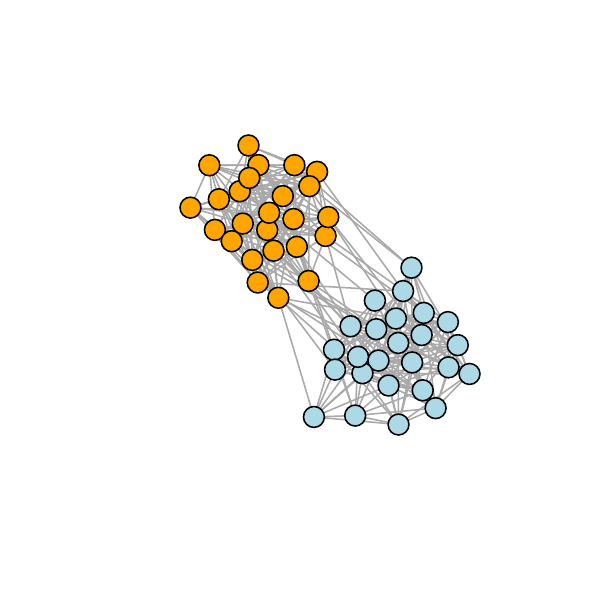} 
        \includegraphics[width=0.3\textwidth, , trim={2cm 2cm 2cm 2cm},clip]{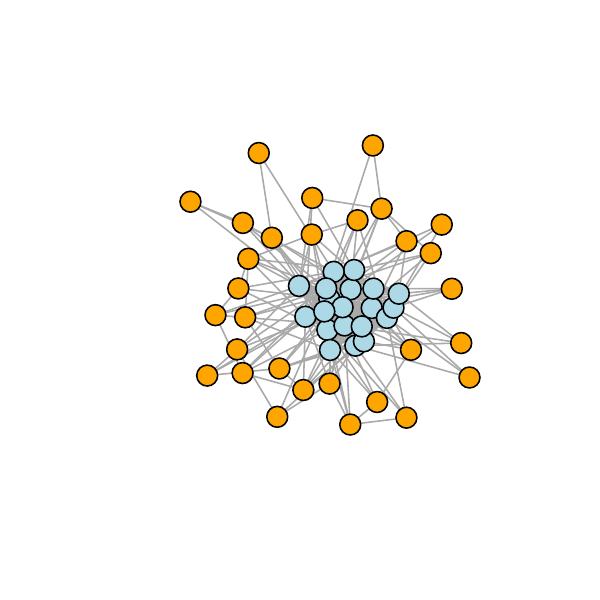} 
     \caption{Community structure (left) versus a core-periphery structure (right).}
    \label{fig:comm_cp}
\end{figure}

\end{itemize}

We now introduce some general mathematical notation for statistical network analysis.
Let $A_{n \times n}$ be the \textit{adjacency matrix} of an undirected network with $n$ nodes.
Then $A$ is a symmetric, binary random matrix with 
$A_{ij} = A_{ji} = 1$ if nodes $i$ and $j$ are connected by an edge, and $A_{ij} = A_{ji} = 0$ otherwise, for $1 \le i<j \le n$.
This is modeled as 
$
A_{ij} \sim \text{Bernoulli}(P_{ij}),
\label{eq-rgm}
$
where $P = \mathbb{E}[A]$ is the symmetric matrix of edge probabilities.
We will write $A \sim P$ as shorthand.
Under this general setting, five central inference problems have been studied in the statistical literature on network data, namely: (i) model development and parameter estimation,   (ii) community detection, (iii) hypothesis testing, (iv) network monitoring, and (v) change detection
and 
model selection.
In the following subsections, we briefly summarize the literature on these five topics.


\subsection{Statistical models and parameter estimation}
   Starting from the classical Erd{ő}s-R{é}nyi (ER) model, where $P_{ij} = p$ for all $i,j$,
the richness of statistical network models has evolved substantially over time.
We briefly describe six well-known models for network data and the corresponding estimators.
In the interest of space, we skip the technical details on rates of convergence.

\begin{itemize}

	\item \textbf{Stochastic Blockmodel (SBM):}
 The stochastic blockmodel \citep{holland1983stochastic,fienberg1985statistical} is probably the most well-studied network model in the statistics literature.
	Under an SBM with $K$ communities,
	$$
	P(i,j) = \omega_{c_i} \omega_{c_j},
	$$
	where $\omega$ is a $K$-by-$K$ symmetric matrix of community interaction probabilities, and ${c} = \{c_i\}_{i=1}^n$ are the communities of the nodes, with $c_i$ taking its value in $1, \ldots, K$ \citep{holland1983stochastic}.
 A number of community detection methods \cite{rohe2011spectral,sengupta2015spectral,zhao2012consistency,gao2017achieving} can be used for estimating the communities $\{\hat{c}_i\}_{i=1}^n$.
 The model parameters $\omega_{rs}$ are estimated as
	$
	\hat{\omega}_{rs} = \frac{\sum_{i,j: \hat{c}_i = \hat{c}_j = r} A(i,j)}{{\hat{n}_r(\hat{n}_r-1)}}${ when } $r=s$, { and } 
	$\hat{\omega}_{rs} = \frac{\sum_{i,j: \hat{c}_i, \hat{c}_j = r} A(i,j)}{\hat{n}_r \hat{n}_s}$ { when } $r \neq s$.
	Here $\hat{n}_r$ is the size of the estimated $r^{th}$ community.
		We estimate $P$ as
		$\hat{P}(i,j) = \hat{\omega}_{\hat{c}_i\hat{c}_j}.$
	
	\item \textbf{Chung-Lu (CL) model:}
	Here
	$$	P(i,j) = \theta_i \theta_j,
 $$
	where $\{\theta_i\}_{i=1}^n$ are the degree parameters \citep{chung2002average}.
	We estimate $P$ as
	$\hat{P}(i,j) = \frac{d_id_j}{2m},$
	where $d_i$ is the degree of the $i^{th}$ node and $m=\sum_{i>j} A(i,j)$. 

 \item \textbf{Random dot product graph (RDPG) model:}
 The RDPG model is a generative model for graphs that assumes the existence of latent positions associated with each vertex, with edge probabilities determined by the dot product of these latent positions \cite{young2007random}. Under the RDPG, with dimension $d$, we have
	${P} = XX',$
	where $X_{n \times d}$ is a matrix of rank $d$ such that $[XX'](i,j) \in (0,1)$ for all pairs $(i,j)$.
  Consistent parameter estimation can be achieved under certain conditions on the sparsity and the rank of the true latent position matrix \cite{athreya2017statistical}. Specifically, it has been shown that the adjacency spectral embedding (ASE) provides consistent estimates of the latent positions as the number of vertices grows \cite{sussman2012consistent}.
	The ASE estimator is  given by 
	$$
 \hat{X} = U_A S_A^{1/2}, 
 $$
	where $S_A$ is the diagonal matrix of the $d$ largest eigenvalues of $(A'A)^{1/2}$ and $U_A$ is the 
 	$n$-by-$d$ 
	matrix 
 	whose columns consist 
	of the corresponding eigenvectors \citep{sussman2012consistent}.
	We estimate $P$ as
$	\hat{P} = \hat{X}\hat{X}'.$
		
	\item \textbf{Latent Space Model (LSM):}
 In the LSM, each vertex is associated with a latent position in a Euclidean space.
 Edge probabilities are determined by $l_2$ distances between latent positions of the nodes, given by
$$
\text{logit}({P}({i,j}))=\alpha - |{z}_i-{z}_j|
$$

for $1\leq i<j\leq n$, where ${z}_i \in \mathbb{R}^d$ is the latent position of the $i^{th}$ and $\alpha$ is a parameter that controls overall sparsity.
  For LSMs, consistency of parameter estimation is generally guaranteed when the latent positions are identifiable and when certain regularity conditions regarding the distribution of latent positions and the link function are satisfied \cite{handcock2007model}.
One can estimate $P$ by using the maximum likelihood estimation strategy described in  \cite{hoff2002latent} as implemented in the R package \textit{latentnet} \citep{krivitsky2008fitting}.

		\item \textbf{Degree Corrected Blockmodel (DCBM):} 
	Under the DCBM with $K$ communities,
	$P(i,j) = \theta_i\omega_{c_i} \omega_{c_j}\theta_j,$
	where $\omega$ is a $K$-by-$K$ symmetric matrix of community-community interaction probabilities, and $\{c_i\}_{i=1}^n$ are the communities of the nodes, and $\{\theta_i\}_{i=1}^n$ are degree parameters \citep{karrer2011stochastic}.
	Several community detection methods \citep{qin2013regularized,sengupta2015spectral, zhao2012consistency,gao2018community} can be used for estimating the communities $\{\hat{c}_i\}_{i=1}^n$.
 The remaining parameters are estimated as
	$\hat{\omega}_{rs} = \sum_{i,j: \hat{c}_i, \hat{c}_j = r} A(i,j)$, 
 { and } $\hat{\theta}_i = \frac{d_i}{\delta_r},$
	{ where } $\delta_r = \sum_{i: \hat{c}_i=r} d_i$
	 is the degree of the estimated $r^{th}$ community.
	We estimate $P$ as
	$\hat{P}(i,j) = \hat{\theta}_i\hat{\omega}_{\hat{c}_i \hat{c}_j} \hat{\theta}_j.$
	\item \textbf{Popularity Adjusted Blockmodel (PABM):} 
	Under the PABM with $K$ communities,
	\[
	P(i,j) = \theta_{ic_j} \theta_{jc_i},
	\]
	where $\theta_{ir}$ represents the popularity of the $i^{th}$ node in the $r^{th}$ community, and $\{c_i\}_{i=1}^n$ are the node communities \cite{senguptapabm}.
	One can use the extreme points method of \cite{le2016optimization} or the sparse subspace clustering of \citep{pensky2019spectral} to estimate communities, and estimate the popularity parameters as
	$\hat{\theta}_{ir} = \frac{\sum_{j: \hat{c}_j=r} A(i,j)}{\sqrt{\sum_{i,j: \hat{c}_i, \hat{c}_j = r} A(i,j)}}.$
\end{itemize}

Furthermore, the exponential random graph models (ERGM) \citep{schweinberger2020exponential}
were proposed to describe a network through a set of network statistics.
There has also been substantial work on \textit{graphons}
\citep{caron2017sparse},
which are symmetric measurable functions that represent every convergent
limit of a network sequence.

\subsection{Community detection}
One of the main inferential tasks on a network with an underlying community structure is to discover the community membership of each node.  
Over the last decade or so, a methodologically versatile and theoretically deep statistical literature on community detection has emerged.
A number of community detection algorithms have been studied in the literature.
This includes
spectral clustering and its variants that leverage the eigen structure of the network \citep{rohe2011spectral, lei2015consistency, sengupta2015spectral},
likelihood based methods that maximize the model likelihood \citep{bickel2009nonparametric, amini2013, zhao2012consistency, senguptapabm}, as well as
    optimization based methods \citep{le2016optimization}.
    
Given the wide variety of community detection methods that have been studied in the literature, a description of all such methods is beyond the scope of this article.
Instead, we provide brief descriptions of three popular community detection methods based on spectral graph theory.

\begin{itemize}
    \item \textbf{Spectral clustering:}
    For spectral clustering, we compute 
the $K$ orthonormal eigenvectors corresponding to the $K$ largest (in absolute value) eigenvalues of the adjacency matrix $A$, and put them in an $\times K$ matrix. K-means clustering is applied on the matrix rows to estimate the node communities \citep{rohe2011spectral}.

\item \textbf{Bias-adjusted spectral clustering (BASC):}
BASC has been recently proposed by \citep{lei2019}.
 Let $D$ be the diagonal matrix of node degrees in $A$.
Here, we use the $K$ dominant eigenvectors of the ``bias-adjusted'' matrix $A^TA-D$ instead of $A$, and apply $K$-means clustering to find communities.

\item \textbf{Spherical $K$-median spectral clustering (MedSC):}
MedSC, as proposed in \citep{lei2015consistency}, has a mechanism comparable to SC, where we compute the $K$ dominant eigenvectors of $A$, and put them in an $n\times K$ matrix. But unlike SC, here we first normalize the rows of the matrix with respect to the respective euclidean norms and then apply $K$-median clustering (instead of $K$-means) on the normalized rows to estimate the subsampled nodes' communities.
Note that, it is also possible to use $K$-means clustering on the normalized rows, as done in \citep{lyzinski2014perfect}.
However, \citep{lei2015consistency} and \citep{lyzinski2014perfect} considered two somewhat different parametric configurations to define the DCBM framework.

\end{itemize}

An issue that appears in the theoretical analysis of community detection methods is the lack of identifiability of community (or cluster) labels.
We now introduce some notation to deal with this issue.
Suppose ${c} = \{c_i\}_{i=1}^n$ are the true communities of the nodes, with $c_i$ taking its value in $1, \ldots, K$.
Any candidate assignment $e = (e_1, \ldots, e_n)$ is an $n$-vector comprising the $K$ community labels where each label appears at least once.
Let $\Pi$ be the symmetric group of all permutations of $\{1, \ldots, K\}$.
For $\sigma \in \Pi$, we define $\sigma(e) := (\sigma(e_1), \ldots, \sigma(e_n))$ as the label permutation of $e$ generated by $\sigma$, and define
$
\Pi(e) = \{\sigma(e):\sigma \in \Pi \}
$
as the set of all label permutations of the assignment vector $e$.
Note that for any $e' \in \Pi(e)$, the community assignments $e'$ and $e$ are identical with respect to community detection.
To address this issue, we define the error rate of a candidate assignment as the proportion of nodes where the candidate assignment and the true assignment disagree, i.e.,
\begin{equation*}
\xi_n(e) = \min_{e' \in \Pi(e)} \frac{1}{n} \sum_{i=1}^{n} I[e'_i \neq c_i],
\end{equation*}
where disagreement is minimized over all label permutations of the candidate assignment.
Now, consider the estimated communities $\hat{c}_i$ from some community detection method.
The community detection error is then defined as $\xi_n(\hat{c})$.
Two notions of community detection consistency have been studied in the literature: weak consistency, where
\[ \xi_n (\hat{c}) \stackrel{P}{\rightarrow} 0, \]
and strong consistency, where
\[ Prob[\xi_n (\hat{c}) \neq 0] \rightarrow 0. \]
It is well-established that spectral clustering and BASC are strongly consistent under the SBM while Med-SC is consistent under the DCBM \citep{lei2015consistency, lei2019}.
    Rate-optimal community detection methods under the SBM and the DCBM have been proposed by \cite{gao2017achieving} and \cite{gao2018community}, respectively.
    Both papers point out that applying a clustering technique as an initialization step, followed by a refinement step, can lead to rate-optimal community detection as long as the initial clustering satisfies certain weak consistency conditions.
    We skip technical details in the interest of space.

    \subsection{Hypothesis testing}
    Consider a set of $n$ entities labeled as $1, \ldots, n$, and two  undirected networks (with no multiple edges or self-loops) representing interactions between them.
This is represented by two $n$-by-$n$ symmetric adjacency matrices $A_1$ and $A_2$, where $A_1(i,j) = 1$ if entities $i$ and $j$ interact in the first network and $A_1(i,j) = 0$ otherwise, and similarly for $A_2$.
The statistical model is given by $A_1 \sim P_1$ (and $A_2 \sim P_2$), which is short-hand for $A_1(i,j) \sim \text{Bernoulli}(P_1(i,j))$ for $1 \le i < j \le n$.
Let $\tau(P)$ represent a network feature of interest, i.e., a function of the probability matrix and the model parameter of interest.
The inference task of interest is to test whether the networks are similar in terms of the feature quantified by $\tau(\cdot)$, i.e., whether $\tau(P_1) = \tau(P_2)$.

This general setting covers a range of problems, depending on the definition of $\tau$.
Two special cases have been studied in the literature: equality and scaling.
Under the equality problem, we want to test whether $P_1 = P_2$,
akin to the classical paired sample testing problem.
However, equality might be too restrictive as a notion of similarity.
For example, consider the Aarhus network representing five pairwise interactions (coauthor, leisure, work, lunch, and Facebook) between  $61$ researchers \cite{rossi2015towards}.
Due to the fundamentally different frequencies of these activities, even if two interactions follow similar patterns, one might occur more commonly than the other (e.g., lunch vs. Facebook).
This notion of similarity can be tested via the \textit{scaling} problem, i.e., testing whether $P_1 = cP_2$ for some $c>0$.
Note that equality is a special case (more restrictive version) of scaling.
The equality problem
corresponds to $\tau(P) = P$ and the scaling problem corresponds to $\tau(P) = P/||P||_F$ where $||\cdot||_F$ represents the Frobenius norm of a matrix.

We call this problem the \textit{matched} network inference problem
to emphasize the fact that the nodes in the two networks are matched, i.e., the $i^{th}$ row of $A_1$ and the $i^{th}$ row of $A_2$ represent the same entity.
This helps us distinguish this problem from three closely related problems.
First, one could have several networks from both distributions, i.e., $A_1, \ldots, A_{m_1} \sim P_A$ and $B_1, \ldots, B_{m_2} \sim P_B$, which is closer to the classical two-sample problem, studied in \cite{ginestet2017hypothesis}.
Second, one could have two networks $A_1 \sim P_1$ and $A_2 \sim P_2$ on possibly non-matching sets of nodes, e.g., legislative co-sponsorship networks from two different US Congresses \cite{fowler2006legislative}.
Third, we may not know which node in $A_1$ corresponds to which node in $A_2$,
which is the well-studied network alignment problem \citep{emmert2016fifty}.

While matched network inference is a relatively new area, some exciting progress has been made in recent years \cite{tang2017semiparametric,tang2017nonparametric,ghoshdastidar2017two, ghoshdastidar2017,ghoshdastidar2018practical,li2018two,agterberg2020nonparametric}.
In \cite{tang2017semiparametric}, \cite{tang2017nonparametric}, and \cite{agterberg2020nonparametric}, the authors studied the problem under the random dot product graph framework and its generalization.
In \cite{li2018two}, the authors studied the problem under the stochastic blockmodel.
In \cite{ghoshdastidar2018practical}, the authors proposed a test based on the spectral norm, and in \citep{ghoshdastidar2017two, ghoshdastidar2017} the authors studied the problem from an information theoretic perspective to derive minimax bounds.
In \cite{tang2017semiparametric}, the authors studied the problem under the RDPG model, proposing 
\begin{equation}
T_{ase}(A_1, A_2) = \min_{W \in \mathcal{O}_n}||\hat{X}_1-\hat{X}_2W||_F,
\label{eq-tase}
\end{equation}
as the test statistic.
Here $\hat{X}_1$ and $\hat{X}_2$ are estimated from $A_1$ and $A_2$ by using ASE and $\mathcal{O}_n$ is the set of all $n$-by-$n$ orthogonal matrices.
Two independent sets of parametric bootstrap iterations are carried out, using the generative models $\hat{X}_1\hat{X}_1^T$ and $\hat{X}_2\hat{X}_2^T$, and the p-value is defined as the larger of the two bootstrap p-values.

In \cite{ghoshdastidar2018practical}, the authors studied the problem under the generic inhomogeneous Erd\"{o}s-Ren\'{y}i model.
Given $A_1 \sim P_1$ and $A_2 \sim P_2$, 
they propose estimating $\hat{P}_1$ and $\hat{P}_2$ as stochastic blockmodel approximations with $r$ communities.
Then, they consider the differenced adjacency matrix, $C$, and construct a scaled version given by
\[
\Tilde{C}(i,j) = \frac{A_1(i,j) - A_2(i,j)}
{\sqrt{(n-1)\left(\hat{P}_1(i,j) (1-\hat{P}_1(i,j) +
\hat{P}_2(i,j) (1-\hat{P}_2(i,j) \right)}}.
\]
The test statistic is given by
\begin{equation}
T_{eig}(A_1, A_2) = n^{2/3}(||\Tilde{C}|| -2),
\label{eq-teig}
\end{equation}
where $||\cdot||$ denotes the spectral norm of a matrix.
The test is rejected for large values of $T_{eig}$ with p-values computed from the
standard Tracy-Widom distribution \citep{tracy1996orthogonal}.
The $T_{eig}$ test has two advantages: first, it is computationally efficient since no bootstrapping is needed, and second, it is highly versatile as it does not require model specification.

There are several important open problems in this research area.
The present literature encompasses two matched networks. There's potential to scale this up to consider multiple matched networks simultaneously.
Another important next step is to consider more intricate notions of similarity, moving beyond equality and scaling, and
extending to other important network features in scientific domains,
e.g., epidemic thresholds \citep{leitch2019toward, dasgupta2022scalable}, and networks with node/edge covariates.

    \subsection{Model selection}
    Model selection is the problem of determining the most appropriate statistical model for a given network.  
    All the network inference methodologies discusssed so far are dependent on the knowledge of the underlying network model. 
    When multiple models might fit the observed network to some extent, model selection helps in identifying the most parsimonious and predictive model, which best explains the observed data.
For example, any community detection algorithm for blockmodels require the knowledge of the true number of communities. Using a smaller or higher number of communities than the truth results in underfitting or overfitting of the network model, thus reducing the overall prediction accuracy. Another model selection problem for blockmodels is testing for the presence of degree heterogeneity in the network. Similarly, any inference on an RDPG requires the knowledge of the latent dimension, $d$. 

A number of methods have been proposed in the literature for estimating the number of communities under the SBM and DCBM. \citet{le2015estimating} proposed a method for estimating the number of communities based on the spectral properties of certain graph operators, such as the non-backtracking matrix and the Bethe Hessian matrix. \citet{lei2016goodness} developed a goodness-of-fit test for the stochastic block model. The test statistic is based on the largest singular value of a residual matrix obtained by subtracting the estimated block mean effect from the adjacency matrix. \citet{bickel2015hypothesis} proposed a hypothesis test based recursive bipartitioning algorithm to estimate the number of communities in an SBM. \citet{ma2021determining} developed a pseudo likelihood ratio statistic to estimate the number of communities in degree-corrected stochastic block models. \citet{hu2020corrected} proposed a corrected Bayesian
information criterion (CBIC) to determine the number of communities in an SBM. 
In a recent pre-print, \citet{cerqueira2023consistent} proved strong consistency of penalized marginal likelihood estimator of the number of communities of a DCBM.
    
In other areas of statistics, cross-validation based methods have proved highly successful and popular in model selection or tuning parameter selection problems.
Two notable papers using the cross-validation approach to model selection are \citet{chen2018network} and \citet{li2020network}.
In \cite{chen2018network}, the authors developed a $V$-fold network cross-validation for determining the number of communities in a blockmodel. 
In \cite{li2020network}, the authors proposed a cross-validation algorithm called Edge Cross Validation ({ECV}) by combining edge sampling and matrix completion. 
Notably, ECV is applicable to a wide array of network model selection problems, including model-free rank estimations, model selection for blockmodels, tuning parameter selection for graphon estimation, dimension selection under RDPG, etc.

    \subsection{Network monitoring and change detection}
Most research in statistical network analysis so far has been focused on static
networks, in which either a single snapshot or aggregated historical
data of a system is used for modeling and analysis. 
However, most
real-world systems are {\textit{dynamic networks}} that exhibit intrinsic time-evolving
behavior, i.e., the underlying structure evolves/changes over time. 
In scientific applications, the occurrence of sudden large changes and shocks
in a network is an important phenomenon
\citep{stevens2021interdisciplinary,stevens2021broader}.
For example, significant changes in the brain connectome network
\citep{xia2013brainnet} can indicate the onset of
a neurological disorder like Alzheimer's disease or epilepsy. {Monitoring}
and {change detection} are therefore crucial for making effective decisions
in a timely manner. Moreover, abrupt changes often affect a network
locally, where only a subset of nodes and their corresponding links
are significantly altered by an event. Consequently, {diagnosis}, defined as identifying
affected sub-networks, plays an important role in root-cause determination
and action planning. For example, it is crucial to determine the parts
of the brain involved in a particular disease, by identifying the
subset of nodes that caused the change in the brain network.

The well-developed field of statistical process monitoring provides a valuable foundational framework for network monitoring \citep{woodall2017overview, jeske2018statistical, stevens2021foundations}.
Network monitoring typically involves the classical two-phase process monitoring framework applied to network data \citep{woodall2014some,woodall2017overview, zhao2018performance,zhao2018aggregation,kodali2019value}.
In Phase I, the user collects a sample of time-varying networks that represent the in-control state, and uses this sample to gain an understanding of the in-control behavior.
In Phase II, the user observes networks successively over time, and the goal is to determine whether there is a significant deviation from the in-control state.
To make this determination at a given point of time, the user is only allowed to use the Phase I sample and the Phase II data collected until that specific time point.
On the other hand, `network change point detection typically involves the scenario where there is no Phase I and Phase II like traditional statistical process monitoring.
Instead, the entire stream of network data is available to the user, and the goal is to retrospectively estimate time points when the model changed.
This problem definition that has been used in several recent statistical papers on network change point detection, such as \citep{bhattacharjee2018change,wang2021optimal}. 
On a related note, it is important to distinguish between \textit{monitoring {of} network data} and \textit{monitoring of data over a network} \citep{sengupta2018discussion}.
The first case arises when we have time series of networks and the network structure itself, i.e., nodes and edges, is the time series variable of interest, e.g., social networks.
The second case arises when the network itself is fixed over time, and we observe certain node-level or edge-level variables, and we are interested in monitoring these variables over time, e.g., the transmission of electric power via the power grid network.

In existing work, network snapshots are typically summarized
into a few measures and statistics which are used for network monitoring, e.g.,  CUSUM and EWMA control charts \citep{mcculloh2011detecting} or
scan statistics \citep{priebe2005scan,marchette2012scan,neil2013scan,park2013anomaly}.
which are used for network monitoring.
In \citep{mcculloh2011detecting}, the authors used cumulative sum
(CUSUM) and exponentially weighted moving average (EWMA) control charts
to monitor graph measures (e.g., density, average closeness, etc.)
calculated from networks. In \citep{priebe2005scan,marchette2012scan,neil2013scan},
the authors proposed a scan statistics approach, and in \citep{park2013anomaly}, the authors  used a fusion of network
statistics (including the scan statistic) to detect changes in a stream
of networks. 

There are several important directions for future research.
In most of the existing literature, network snapshots are summarized
into a few measures and statistics, which are used for network monitoring.
This may cause information loss in networks that affects the detection
power of the monitoring approach. 
Moreover, most existing methods are not robust to intrinsic slow
changes of time-varying networks, which results in a high false alarm
rate. 
Finally, very little research can be found on diagnosis in time-varying
networks.

\section{Future: emerging research directions}
Statistical network analysis is a highly dynamic area of research, and compelling new trends are constantly emerging.
This section focuses on two emerging research directions that are aimed at addressing two barriers that impede effective statistical analysis of networks:  computational scalability and hypothesis testing for network phenomena.
The first barrier stems from the computational limitations of existing methods, 
making it infeasible 
to carry out statistical inference tasks on large-scale networks.
The second barrier prevents the rich set of statistical models from being utilized in several network problems where current methods are too simplistic to be used beyond the simplest homogeneous models.

\subsection{Scalable network inference}
While existing statistical methods for network data inference are statistically sound, with rigorous theoretical guarantees, they are not computationally feasible for networks larger than a few thousand nodes.
Furthermore, since most existing methods do not use distributed storage or parallel processing, they cannot utilize these contemporary computing architectures to improve computational performance.
This makes the statistical toolbox of community detection methods impractical in many scientific fields where massive network data are becoming increasingly common, e.g., epidemic modeling \citep{venkatramanan2021forecasting}, online social networks \citep{guo2020online, leskovec2012learning}, and biomedical text networks \citep{komolafe2022scalable}.

How serious is this problem?  To illustrate this, we report a brief computational experiment involving spectral clustering \citep{rohe2011spectral}, a popular and relatively fast community detection method.
We generated networks from the SBM with $(n,K)= (100000, 20)$ and $(150000, 30)$.
In Table \ref{memcomp}, we report three aspects of the computational cost from this study: 
(i) data storage size, 
(ii) the memory requirement (given by \textit{peakRAM}) for creating/ accessing the intermediate matrices involved in spectral clustering, and 
(iii) the runtime of the algorithm.
We observe that, even for such moderately sized networks, the memory requirement is well over the 8000 Mb RAM of a typical laptop and the runtime is as high as 20-40 minutes.
Note that spectral clustering is one of the fastest community detection algorithms \citep{mukherjee2021two, wang2021fast}, especially with our implementation in R using fast spectral algorithms from the package \textit{irlba}.
Other community detection algorithms, e.g., likelihood-based methods, are likely to fare much worse.

\begin{table}
\centering
\small
\begin{tabular}{|c|c|c|c|c|}
   \hline
   $n$ & $K$ & storage size (Mb.) & memory requirement (Mb.) & runtime (seconds)\\
  \hline
   100000 & 20 & 120.0 & 9232.5 & 1126.7\\ 
   150000 & 30 & 270.0 & 20934.5 & 2439.0 \\
   \hline
\end{tabular}
\caption{Computational cost of spectral clustering, with units in parentheses, on a server with Intel Xeon(R) E5-4627 v3 processors.
The R functions \textit{irlba} and \textit{peakRAM} were used to implement spectral clustering and to compute the memory requirement, respectively.}
\label{memcomp}
\end{table}

Classical strategies such as {subsampling} 
\citep{ma2015statistical,politis1999subsampling}
and 
 {divide-and-conquer}
\citep{kleiner2014scalable,jordan2019communication},
that have worked well in other areas of massive data inference,
unfortunately do not work for
networks due to two reasons.
First, each node has its own parameter, which rules out subsampling
since each node must be included. 
For example, 
Consider the RDPG model, where $P = XX^T$ and the $i^{th}$ row of $X_{n \times d}$ represents the latent position of the $i^{th}$ node \citep{young2007random}.
if the $i^{th}$ node is not in the subsample, then the subsample carries no information about the $i^{th}$ row of $X$, making it impossible to estimate $X$.
Second, the model parameters are not identifiable due to invariance properties, e.g., under the RDPG model, $P = XX^T = (XQ)(XQ)^T$ for any orthogonal matrix $Q_{d \times d}$, which means $X$ and $XQ$ are invariant.
Therefore, results from disjoint data subsets cannot be aggregated, and divide-and-conquer does not work.
To see this, suppose we partition 
$X = \begin{pmatrix}
    X_1\\X_2
\end{pmatrix}$ and 
$P = XX^T = 
\begin{pmatrix}
    X_1X_1^T \; X_1X_2^T \\
    X_2X_1^T \; X_2X_2^T
\end{pmatrix}$,
and 
estimate $X_1$ and $X_2$ separately.
Due to invariance, we end up with estimates of $X_1Q_1$ and $X_2Q_2$, where $Q_1$ and $Q_2$ could be any $d\times d$ orthogonal matrices.
Since $X_1Q_1 (X_2Q_2)^T \neq X_1 X_2^T$ in general, combining the estimates does not provide a valid estimate of the top-right quadrant of $P$.
Therefore, we need new ideas for carrying out statistical inference tasks on large-scale networks.

As another example, consider a large-scale network with a million nodes belonging to 10 communities (or latent groups), and suppose we are interested in estimating the community membership of each node by using spectral clustering.
Suppose we partition this network into 20 subnetworks (each with $m=50,000$ nodes) and carry out spectral clustering on each subnetwork, in parallel.
For each subnetwork, we have now estimated the community membership for the nodes in that subnetwork.
However, this does not accomplish our goal of estimating the community membership of all nodes in the original network.
This is because community detection has an intrinsic non-identifiability issue \cite{rohe2011spectral,senguptapabm} for community labels.
The first community in subnetwork 1 could correspond to any of the $10$ communities in subnetwork 2, and similarly for the other subnetworks.
This means that there are $10!$ possible combinations to aggregate the community memberships across the first two subnetworks, and there is no way to know which combination is correct.

In related work, \cite{amini2013} developed a pseudo-likelihood approach to address the computational challenge of computing the likelihood under the SBM, and recently this strategy was improved by \cite{wang2021fast}.
Two recent preprints \citep{wu2020distributed,zhang2022distributed} have also studied this problem under the blockmodel framework.
A notable recent contribution is
\cite{mukherjee2021two}, where the authors proposed two scalable methods for community detection based on a divide-and-conquer strategy.
In particular, the methods proposed by \cite{mukherjee2021two} apply to a broad range of community detection methods and models.
In \citep{chakrabarty2023sonnet}, the authors proposed another general-purpose subsampling based divide-and-conquer algorithm, SONNET, for community detection in large networks. The
algorithm splits the original network into multiple subnetworks with a common
overlap, and carries out detection algorithm for each subnetwork. The results
from individual subnetworks are aggregated using a label matching method to
get the final community labels.

So far, work on scalable network inference has been mostly focused on community detection. 
An important direction of future research is to develop scalable methods for other inference problems, such as parameter estimation, hypothesis testing, model selection, and network monitoring.
Furthermore, most existing methods work on a specific model or method, which makes them less automatic for practitioners as they have to use different algorithms for each inference task/method/model.
Developing generalizable techniques that can be used for various models and inference tasks is an important research direction.

\subsection{Hypothesis testing for network phenomena}

While some aspects of network data (e.g., community structure) have been extensively studied in the statistics literature, many other aspects, often of similar or greater relevance to network science, have received less attention from the statistics research community.
In the absence of formal statistical frameworks, heuristic metrics are used in quantifying the strength of these properties. 
     These metrics are often compared to simplistic benchmark models such as the homogeneous Erd{ő}s-R{é}nyi (ER) model.
For example, a given network is deemed to exhibit small-world property or community structure if the metric exceeds benchmarks based on the ER model.
This has led to an undesirable situation where existing methods have little consideration of the uncertainty and statistical significance of the metric being used.

\begin{figure}
\centering
  \centering
	\includegraphics[height=0.15\textheight]{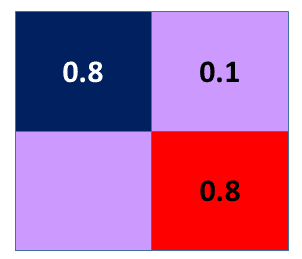}
	\hspace{1ex}
	\includegraphics[height=0.15\textheight]{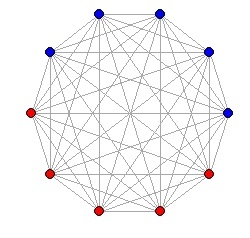}\\
	\vspace{1ex}
	\includegraphics[height=0.15\textheight]{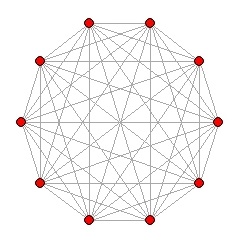}
	\hspace{2ex}
	\includegraphics[height=0.15\textheight]{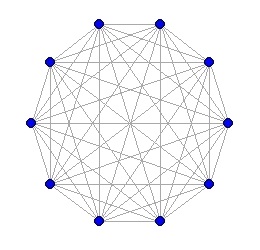}
	\caption{SBM (top left) with two communities and 50 nodes. All cliques have 10 nodes, but the ``mixed'' clique (top right) is anomalous while the ``pure'' cliques (bottom) are very likely to appear randomly.
 \label{fig:clique_eg}}
\end{figure}

    Consider the anomalous clique detection problem, where
we want to determine whether a given network contains an {anomalous} clique.
A clique is a fully connected subgraph where all node pairs are connected (Fig. \ref{fig:clique_eg}).
Note that cliques can form randomly, e.g., a network generated from the ER model contains a clique with $2 \log_{{1}/{p}}(n)$ nodes almost surely \citep{bollobas1976cliques}.
 %
 The goal here is to detect \textit{anomalous} cliques that are very unlikely to appear randomly.
 This is 
a famously well-studied problem 
 under the homogeneous ER model
\citep{alon1998finding,deshpande2015finding}.
However, the existing methods are based on the premise that
in order to determine whether a clique is anomalous, we only need to consider the size of the clique.
 While this premise is true for the ER model, it does not carry over to even the simplest inhomogeneous models.
To illustrate this, 
10000 networks were generated from a toy SBM (Fig. \ref{fig:clique_eg}, top left) with two communities (red and blue) of 25 nodes each,
where intra-community edges and inter-community edges
appear with a probability of 0.8 and 0.1, respectively.
 None of these graphs contained a  10-node ``mixed'' clique consisting of five red nodes and five blue nodes (Fig. \ref{fig:clique_eg}, top right), but 87\% of them contained at 
 least one 10-node ``pure'' clique consisting entirely of red nodes or blue nodes (Fig. \ref{fig:clique_eg}, bottom).
 A 10-node mixed clique is, therefore, highly anomalous for this benchmark model, but a 10-node pure clique is not. 
 Even in this simple model, it is impossible to determine whether a clique is anomalous by only looking at its size.
 Thus, the very premise of the existing methodology
 fails to generalize to statistical models beyond the ER model.

 Two other related threads in the literature are on using scan statistics and spectral methods for anomalous clique detection.
    The scan statistic approach has been well-studied from a theoretical perspective \citep{arias2014community}, including recent work under the Chung-Lu model by \citep{bogerd2020cliques}. However, the literature is purely theoretical as this method requires scanning over all (or nearly all) subgraphs, which is impossible in finite time, even for small networks (say, $n=50$).
 Heuristic spectral methods have also been studied \cite{miller2015spectral}, but without any theoretical justification, and \citet{komolafe2019statistical} showed that the heuristics are questionable.
 Taken together, there is no finite-time method for anomalous clique detection that is statistically valid under non-ER models.
 This is an important open problem that can greatly benefit from methodological work by the statistics community.

The generalizability bottleneck affects several other areas of network analysis.
For example, consider the celebrated small-world property, where a network simultaneously exhibits \textit{segregation}, i.e., its nodes form small and tightly knit groups scattered across an otherwise sparse network, and \textit{integration}, i.e., on average, two nodes are at a small distance \citep{watts1998collective}. 
Recall that the small-world coefficient is defined as $\sigma = \frac{{C}/C_{ER}}{{L}/L_{ER}}$, 
where $C_{ER}$ and $L_{ER}$ are the expected values of ${C}$ and ${L}$ in an ER model of same density as the observed network \citep{watts1998collective,humphries2005brainstem,guye2010graph}.
The current approach is to consider a network to be small-world if $\sigma>1$.
There are several shortcomings of this approach.
The coefficient $\sigma$ is a purely empirical metric that 
does not account for the randomness of $A$ \cite{telesford2011ubiquity,hilgetag2016brain}, and uses the simplistic ER model as a ``hard-coded'' benchmark.
Any network with ${C}$ sufficiently larger than the ER model is deemed to be small-world under the current framework, 
leading to a false sense that the small-world property is near-universal.
In related work, \citet{lovekar2021testing} recently proposed a formal inference framework for the small-world property by developing an intersection test that decouples the properties of high transitivity and low average path length as separate events to test for. 
This is another exciting and important research area where the statistical community can contribute by developing formal statistical frameworks for uncertainty quantification and hypothesis of network phenomena, such as core-periphery structures, small-world property, etc.

\section{Discussion}

This review article briefly charted the past, present, and future of statistical network analysis.
The aim was to make two connections: (i) how the key historical developments laid the foundation for today's statistical network analysis
and (ii) how statistical network analysis is a central component of the wider discipline of network science.
Given the breadth and depth of the literature, many important topics could not be covered in this article.
This includes multilayer networks, dynamic networks, Bayesian inference on networks, the geometry of the latent space of network models, and higher-order interactions, to name a few.

In conclusion, it is clear that the statistical analysis of networks is not just a field marked by methodological innovation and scientific significance but one that is also in a state of rapid and exhilarating evolution. The potential directions for future research are as diverse as they are promising, appealing to statistical researchers to work on important and exciting problems that promise to advance this field further. 

\bibliographystyle{apalike}
\bibliography{ref}

\end{document}